\documentclass[conference]{IEEEtran}
\IEEEoverridecommandlockouts
\usepackage{cite}
\usepackage{amsmath,amssymb,amsfonts}
\usepackage{algorithmic}
\usepackage{graphicx}
\usepackage{textcomp}
\usepackage{bm}
\usepackage{xcolor}
\def\BibTeX{{\rm B\kern-.05em{\sc i\kern-.025em b}\kern-.08em
    T\kern-.1667em\lower.7ex\hbox{E}\kern-.125emX}}
\begin{document}

\title{Bayesian Optimization of Area-based Road Pricing}

\author{\IEEEauthorblockN{Renming Liu}
\IEEEauthorblockA{\textit{DTU Management} \\
\textit{Technical University of Denmark}\\
Copenhagen, Denmark \\
liure@dtu.dk}
\and
\IEEEauthorblockN{Yu Jiang}
\IEEEauthorblockA{\textit{DTU Management} \\
\textit{Technical University of Denmark}\\
Copenhagen, Denmark  \\
yujiang@dtu.dk}
\and
\IEEEauthorblockN{Carlos Lima Azevedo}
\IEEEauthorblockA{\textit{DTU Management} \\
\textit{Technical University of Denmark}\\
Copenhagen, Denmark  \\
climaz@dtu.dk}
}
\IEEEoverridecommandlockouts
\IEEEpubid{\makebox[\columnwidth]{978-1-7281-8995-6/21/\$31.00 \copyright~2021 IEEE \hfill} \hspace{\columnsep}\makebox[\columnwidth]{ }}

\maketitle

\vspace{-8pt}

\begin{abstract}
This study presents a Bayesian Optimization framework for area- and distance-based time-of-day pricing (TODP) for urban networks. The road pricing optimization problem can reach high level of complexity depending on the pricing scheme considered, its associated detailed network properties and the affected heterogeneous demand features. We consider heterogeneous travellers with individual-specific trip attributes and departure-time choice parameters together with a Macroscopic Fundamental Diagram (MFD) model for the urban network. Its mathematical formulation is presented and an agent-based simulation framework is constructed as evaluation function for the TODP optimization problem. The latter becomes highly nonlinear and relying on an expensive-to-evaluate objective function. We then present and test a Bayesian Optimization approach to compute different time-of-day pricing schemes by maximizing social welfare. Our proposed method learns the relationship between the prices and welfare within a few iterations and is able to find good solutions even in scenarios with high dimensionality in the decision variables space, setting a path for complexity reduction in more realistic road pricing optimization problems. Furthermore and as expected, the simulation results show that TODP improves the social welfare against the no-pricing case.
\end{abstract}


\section{Introduction}
As one of the most promising traffic demand management policies for mitigating traffic congestion \cite{langmyhr1999understanding}, congestion pricing has been investigated extensively over a century. Various pricing models and solution methods have been proposed in theory and practice \cite{yang2005mathematical,lindsey2006economists,VonKnoordegraaf2014,gu2018congestion}. These are often categorized based on the type of infrastructure they are applied to \cite{de2011traffic}:
\paragraph{Facility-based schemes} Under such schemes tolls are imposed at a single or multiple points on a specific facility such as a road link, bridge, tunnel, or levied on a part of the facility, e.g.: High Occupancy Vehicle (HOV) lanes.

\paragraph{Cordon-based schemes} Travelers crossing a cordon in the inbound and/or outbound directions are charged, while those travelling inside the cordoned area are not.

\paragraph{Area-based schemes} Travelers are required to pay a toll when entering, exiting or traveling within a defined area.

\paragraph{Distance-based schemes} The charges are associated with the trip length linearly or nonlinearly \cite{meng2012optimal}, mainly  to enhance fairness and efficiency compared to the previous purely access-based controls \cite{daganzo2015distance}.

Compared to facility-based schemes, area-based schemes are able to intercept more trips and less sensitive to traffic diversion \cite{olszewski2005modelling}. It could simplify the complexity of the problem in both theory and practice as it avoids link-based pricing and, in some cases, does not require knowing the within-network dynamics, link attributes, and detailed information of path selection for different origins and destinations. Yet these simplifications also encompass limitations in efficiency and applicability compared to more disaggregate controls \cite{de2011traffic}.

Despite the differences in the above schemes, finding an optimal toll value for practice is often associated with a computationally expensive objective function, high dimensional decision variables (due to a large number of possible price values of a step-toll \cite{ROBINLINDSEY201246}, dynamic link-specific tolls for large networks \cite{Joksimovic2005} or realistic demand and supply modelling features \cite{LENTZAKIS2020102685} in the optimization evaluation function). The optimization problems often do not have a closed form solution that can be solved analytically. Alternatively, simulation-based optimization approaches are useful to handle such toll optimization problem with an expensive-to-evaluate objective function and high non-linearity, as these approaches only require the paired data of decision variables and objective function values to search for optimal decision variables \cite{amaran2016simulation, LENTZAKIS2020102685, chen2014surrogate}. In the existing applications of simulation-based optimization on pricing toll scheme design, there are two different approaches: feedback control and surrogate-based optimization. \cite{zheng2012dynamic} proposed a discrete integral controller to adjust the cordon-based time-dependent charge rates so that the accumulation of the network does not exceed the flow-maximizing value. \cite{zheng2016time} further developed a proportional-integral controller to iteratively control area-based toll rates, and the results demonstrated that the later controller outperformed the former one due to higher flexibility. Recently, \cite{LENTZAKIS2020102685} used feature-variant clustering methods for toll-area definition under distance- and area-based schemes.

On the other hand, surrogate-based optimization aims to approximate the map from the decision variables to the objective function values within a few iterations. For example, \cite{chow2014surrogate} constructed the surrogate model using a radial basis function to optimize the link-specific tolls for networks, and showed that the proposed method converged faster than the genetic algorithm. \cite{chen2014surrogate} compared different surrogate models for the link toll optimization problem and found that kriging (or Gaussian process regression) performed best. \cite{gu2019surrogate} applied the surrogate model with expected improvement sampling to optimized the step toll parameters to control the accumulation of the network.


In this study, we exploit the potential of using Bayesian Optimization (BO) for determining the optimum price in distance- and area-based pricing schemes. We focus on the morning commute problem and on the design of time-of-day pricing (TODP) profiles for networks with heterogeneous travelers. TODP is usually preferable rather than flat tolls, as it can internalize the dynamic congestion costs that travelers impose on each other \cite{de2011traffic}, reduce the social costs of trips in non-congested periods and help improving overall network performance \cite{de2004time}. We also extend other area-based pricing optimization frameworks to the objective of social welfare maximization instead of purely network performance measures. We use a Gaussian (mixture) function to parameterize the TODP scheme for the purpose of reducing the dimension of the decision variables. 

Our hypothesis is that BO can handle the stochasticity from relatively complex simulation-based evaluation functions and find good solutions within few evaluations. BO's implicit surrogate-based method plays a role in mapping the decision variables (i.e., the pricing toll scheme) to the objective function (i.e., the social welfare) and its inherited uncertainty from the demand and network simulation models. For the latter, we rely on the recent developments in (network) Macroscopic Fundamental Diagram (MFD) modeling founded by \cite{daganzo2007urban, geroliminis2007macroscopic} for capturing the dynamics of travel speed, travel production and network accumulation under different pricing setups. Furthermore, we employ the \textit{trip-based MFD} formulation proposed in \cite{fosgerau2015congestion,lamotte2016morning,mariotte2017macroscopic} for allowing for the inclusion of heterogeneous trip lengths in both the control feature and the network performance function.

\section{Methodology}\label{section:simulation}

\subsection{Trip-based MFD Model}\label{section:MFDmodel}

As defined in \cite{lamotte2016morning,mariotte2017macroscopic}, the \textit{trip-based MFD} considers the trip distance traveled by traveler $i$ as a function of trip departure time $t_i^{dep}$ and arrival time, $t_i^{dep}+T_{i}(t_i^{dep})$, within the network. It models it as the integration of the network travel speed over the travel time period, i.e.,
\begin{equation}
\label{trip_l}
L_i = \int_{t_i^{dep}}^{t_i^{dep}+T_{i}(t_i^{dep})} V(n(t))dt
\end{equation}
where $L_i$ is the trip length of traveler $i$, $V(\cdot)$ represents the travel speed and $n(t)$ is the accumulation of the network at time $t$. In this work, the area-based network is considered as a single-reservoir in which the traffic congestion is assumed to be distributed in space with small heterogeneity\cite{daganzo2007urban,geroliminis2007macroscopic}. The traditional MFD assumption that the travel speed $V(n(t))$ is the same for all travelers in the network at time $t$ holds. An event-based simulation is then employed to represent the \textit{trip-based MFD} process, as proposed by \cite{mariotte2017macroscopic,yildirimoglu2020demand}:
\vspace{4pt}

\emph{Step 1.} Initialization: Specify $t_i^{dep}$, $L_i$, $V(n)$ and the total number of travelers $N$; set $n=0$, calculate the initially estimated arrival time for all travelers $\forall i , 1...N$ by $L_i/V(0)$ using \eqref{trip_l}. Note that $n(t)$ is simplified as $n$.

\emph{Step 2.} Construct the event list in the order of time, which consists of $N$ departures and $N$ arrivals.

\emph{Step 3.} Calculate the remaining trip distance for travelers who have entered the network and not finished their trip by $L_i=L_i-V(n)\cdot (t_j-t_{j-1})$
 
\emph{Step 4.} For every event, if the next closest event is another traveler $i'$ departure, set $n=n+1$ and remove this event from the event list; otherwise, set $n=n-1$, output the experienced travel time of  $i$, $T_{i}(t_{i}^{dep})$, and remove this event from the list.
 
\emph{Step 5.} Update the current average traveling speed $V(n)$; update the estimated arrival time for the rest of travelers in the network by using eq. \eqref{trip_l}; and sort the event list in the order of time again.
 
\emph{Step 6.} Go to \emph{Step 3} until all travelers have ended their trips.

Note that the heterogeneous trip length can be accommodated by the \textit{trip-based MFD} model as it handles the travel distances of all travelers separately.

\subsection{Day-to-Day Dynamics Under TODP Strategy}

In a morning commute problem, travelers make departure time choices within a predetermined time window based on perceived travel costs \cite{vickrey1969congestion}, often modeled via the discrete choice model \cite{ben1984dynamic}. Let $t$ represent $t_{i}^{dep}$ for simplicity, and $C_{i,d}(t)$ be the observable monetary utility of traveler $i$ departing at time $t$ on day $d$, the probability of choosing departure time $t$ on day $d$, $Pr_{i,d}(t)$, is written as,
\begin{equation}\label{prob}
    Pr_{i,d}(t)= \frac{\exp\big(\mu\cdot C_{i,d}(t)\big)}{\sum_{s\in TW_i}\exp\big(\mu\cdot C_{i,d}(s)\big)}
\end{equation}
where $\mu$ is the scale parameter and $TW_i$ is the time window of traveler $i$. The utility of the departure time choice for traveler $i$ then consists of the observable utility and a random term which captures the unobserved utility, written as follows,
\begin{equation}\label{utility}
    U_{i,d}(t)= C_{i,d}(t)+ \epsilon_i
\end{equation}
where $\epsilon_i$ is assumed to be i.i.d., following the extreme value distribution, Gumbel distribution.

Under a given pricing scheme, the monetary metric utility includes three components, the travel time cost, schedule delay cost (the difference between realized arrival time and desired arrival time) and the pricing toll payment,
\begin{equation}
\begin{aligned}
\label{exp_cost}
    c_{i,d}(t) =& -\theta_i\cdot \Big[T_{i,d}(t)+\delta_i\cdot SDE_i\cdot\big(T_i^*-t-T_{i,d}(t)\big)\\
    &+(1-\delta_i)\cdot SDL_i\cdot\big(t+T_{i,d}(t)-T_i^*\big)\Big]\\
    &- Toll(t)\cdot L_i\cdot w\\
    =& -\theta_i\cdot tc_{i,d}(t)- Toll(t)\cdot L_i\cdot w
\end{aligned}
\end{equation}

The first term represents the sum of the travel time and schedule delay costs, where $\theta_i$ is the value of time for traveler $i$, $T_{i,d}(t)$ is the travel time for traveler $i$ on day $d$ departing at time $t$, $SDE_i$ and $SDL_i$ are the schedule deviation penalty parameters for early and late arrival for traveler $i$, and $\delta_i$ is a binary variable that equals 1 if $i$ arrives early and 0 otherwise. It is worth to note that the simulation in Section \ref{section:MFDmodel} can only measure the travel time for the departure time chosen by traveler $i$. \cite{lamotte2015dynamic} proposed to estimate the travel time for the not-chosen departure time by assuming there are fictional travelers departing at these departure times without being counted when update the number of travelers in the network. The second term represents the trip length-specific pricing toll payment, where $Toll(\cdot)$ is a time-dependent toll function and $w$ is a constant used to scale down the magnitude of the trip length. In this study, the toll function is parameterized by a Gaussian function, $Toll(t|A, \xi, \sigma)=A\times e^{\frac{-(t-\xi)^2}{2\sigma^2}}$, which is controlled by three parameters, mean $\xi$, variance $\sigma$ and amplitude $A$. Without loss of generality, this toll function can be extended to a linear Gaussian mixture function to provide asymmetry and flexibility. This issue will be further investigated with numerical experiments in Section \ref{Numerical_results}.

At the end of each day $d$, travelers update their perception of the observable monetary utility for day $d+1$ by the linear weighted sum of the initially perceived generalized costs on day $d$ and the experienced (chosen alternative) or estimated (unchosen alternatives) utility on day $d$, as follows:
\begin{equation}\label{learning}
    C_{i,d+1}(t)=\omega\cdot C_{i,d}(t)+(1-\omega)\cdot c_{i,d}(t)
\end{equation}
where the weight coefficient $0<\omega<1$ is the learning parameter, which is here assumed to be identical among travelers in this study. This day-to-day evolution process is expected to converge to an equilibrium state that the perceived and the experienced (or estimated) utility remain the same afterwards and there is no significant changes in departure time choices. See Section \ref{section:d2d_converge} for details on the equilibrium process.

It is relevant to note that the aforementioned day-to-day dynamics is essentially a stochastic simulation, of which randomness comes from different sources such as random choice parameters, random trip attributes or even the probabilistic day-to-day departure time decision process.

\section{Simulation-based Optimization Framework}
\subsection{Objective function}
To evaluate the performance of the proposed pricing scheme, we compare the social welfare per capita $W$ at the equilibrium states of scenarios with and without pricing. For the no toll equilibrium (or NTE), combined with \eqref{utility} and \eqref{exp_cost}, the social welfare per capita $W_{NTE}$ is computed as,
\begin{equation}
\begin{aligned}
\label{sw1}
    W_{NTE}=&\frac{1}{N}\sum_{i=1}^N U_{i,*}(t_{i,*}^{dep})\\
    =& \frac{1}{N}\sum_{i=1}^N\Big[-\theta_i\cdot tc_{i,*}(t_{i,*}^{dep})+\epsilon_i(t_{i,*}^{dep})\Big]
\end{aligned}
\end{equation}
where the subscript '$*$' denotes the equilibrium state.

For the scenario with the TODP scheme, the social welfare $W_{TODP}$ is the consumer surplus (CS) plus the regulator revenue (RR) from charging travelers, which can be written as follows,
\begin{equation}
\begin{aligned}
\label{sw3}
    W_{TODP} =& CS+RR\\
    =& \frac{1}{N}\sum_{i=1}^N\Big(c_{i,*}(t_{i,*}^{dep})+\epsilon_i(t_{i,*}^{dep})+Toll(t_{i,*}^{dep})\cdot L_i\cdot w\Big)\\
    =& \frac{1}{N}\sum_{i=1}^N\Big[-\theta_i\cdot tc_{i,*}(t_{i,*}^{dep})+\epsilon_i(t_{i,*}^{dep})\Big]
\end{aligned}
\end{equation}
Note that the welfare computations in both scenarios are equivalent and consist of travel time cost, schedule delay cost and the random component.

\subsection{Bayesian Optimization}

The BO has two core iterative steps \cite{frazier2018tutorial}: 1) Update the Gaussian Process (GP) that provides a posterior probability distribution of the objective function; and 2) Determine where to sample the next decision variables by optimizing a given acquisition function.

\subsubsection{Gaussian Process}
With a mean function $\mu(\bm{x})$ and covariance function (or kernel) $k(\bm{x},\bm{x'})$, where $\bm{x}$ represents the decision variables, which are the parameters of the pricing toll scheme, the GP regression can approximate the objective function using the historical observations, by taking the distribution of objective function values and decision variables to be a multivariate Gaussian distribution.

Suppose we have $m$ observed objective values as $\mathcal{D}_m=\{\bm{x}_{1:m},W_{1:m}\}$, where $\bm{x}_{1:m}=[\bm{x}_1,\bm{x}_2,\dots,\bm{x}_m]^T$ are the decision variables and $W_{1:m}=[W_1,W_2,\dots,W_m]^T$ are the corresponding objective values, and wish to infer the objective function value at some new decision variables $\bm{x}_{m+1}$. We can first derive the joint distribution of $W_{1:m}$ and $W_{m+1}$ as follows, wherein the mean function is $\mu(\bm{\cdot})$ is assumed to be zero function for simplicity \cite{williams2006gaussian}:
\begin{equation}\label{joint_dist}
\left[\begin{array}{c}
W_{1:m}\\
W_{m+1}
\end{array} 
\right]~\sim~\mathcal{N}\Bigg(\bm{0},~\left[\begin{array}{cc}
\bm{K} & \bm{k}\\
\bm{k}^T & k(x_{m+1},x_{m+1})
\end{array} 
\right]\Bigg)
\end{equation}
where $\bm{k}=[k(\bm{x}_{m+1},\bm{x}_1),k(\bm{x}_{m+1},\bm{x}_2),\dots,k(\bm{x}_{m+1},\bm{x}_m)]^T$, and $\bm{K}$ is the covariance matrix with entries $\bm{K}_{i,j}=k(\bm{x}_{i},\bm{x}_j)$ for $i,~j\in \{1,2,\dots,m\}$.

We can then compute the conditional distribution of $W_{m+1}$ using Bayes’ rule:
\begin{equation}\label{posterior}
    W_{m+1}|W_{1:m}~\sim~\mathcal{N}\Big(\mu(\bm{x}_{m+1}),~\sigma^2(\bm{x}_{m+1})\Big)
\end{equation}
where $\mu(\bm{x}_{m+1})=\bm{k}^T\bm{K}^{-1}W_{1:m}$ and $\sigma^2(\bm{x}_{m+1})=k(x_{m+1},x_{m+1})-\bm{k}^T\bm{K}^{-1}\bm{k}$.

The kernel encodes the correlation between two sets of decision variables via a parametric structure, which is required to be a positive semi-definite function \cite{frazier2018tutorial}. In this work, we adopt the commonly used Matern kernel \cite{matern2013spatial}.

\subsubsection{Acquisition Function}

Based on the inferred mean and variance of the objective function values, acquisition function determines the next decision variables with the maximum utility considering the trade-off between exploration and exploitation. This study uses a popular acquisition function, the upper confidence bound (UCB) \cite{srinivas2009gaussian}, as follows:
\begin{equation}\label{ucb}
    \alpha_{UCB}(\bm{x};\beta)=\mu(\bm{x})+\beta\sigma(\bm{x})
\end{equation}
where $\beta$ determines the balance between exploration and exploitation. A larger $\beta$ leads to more exploration. 

\subsubsection{Space-filling Experimental Design}

To enhance the solution quality and optimization efficiency, it is useful to start with initial sets of decision variables that has a large coverage of the feasible region. The Latin Hypercube Sampling (LHS) is used in this study for the initial space-filling design, which can generate independent sampled sets of decision variables without overlap, providing a better representative of the real variability than random generation and Monte Carlo methods.

\subsubsection{High dimensional BO Using Dropout}

Let $D = \text{dim}(\bm{x})$ represent the dimension of the decision variables. When $D$ increases, the size of the search space grows exponentially, the maximization of the acquisition function becomes more complex and the estimation of the GP is less accurate \cite{kandasamy2015high}. All these three issues impact the performance of BO in high dimension. \cite{ijcai2017-291} tackles these challenges by using the \textit{Dropout} strategy from deep neural networks, that is, BO steps are implemented on the subspace of different $d$ out of $D$. Its performance is tested in Section \ref{Numerical_results} via numerical experiments.

\section{Numerical Experiments}\label{Numerical_results}
This section presents the results of (1) the convergence of the day-to-day departure time choices with and without the TODP scheme; (2) the performance of the BO; and (3) the comparison between the optimized TODP against the NTE case. The numerical settings are presented in Table \ref{settings}.

\begin{table}[b]
\vspace{-7pt}
\centering
  \caption{Numerical settings}\label{settings}
  \begin{tabular}{p{0.13\textwidth}l}
\hline\noalign{\smallskip}
Parameters  & Specification \\
\noalign{\smallskip}\hline\noalign{\smallskip}
Demand &  $N=3700$ [traveler]\\
Trip length &  $L_i=4600+\mathcal{N}(0,(0.2*4600)^2)$ [m], $L_i>0$\\
Scale factor of trip length & $w=2\times 10^{-4}$ \\
Value of time & $\theta_i=1.1$ [DKK/min] \cite{fosgerau2007danish}\\
Schedule deviation penalty & $\begin{bmatrix} SDE_i \\ SDL_i \end{bmatrix}=\begin{bmatrix} 0.5\\ 4 \end{bmatrix}+\mathcal{N}\Bigg(\begin{bmatrix} 0.05^2 & 0.1^2\\ 0.1^2 & 0.4^2 \end{bmatrix}\Bigg)$\\
& $SDE_i \in [0.3,0.7]$, $SDL_i \in [2.5,5.5]$\\
Network capacity & $n_{jam}=4500$ [vehicle] \\
Free flow speed & $v_f=9.78$ [m/s] \\
Speed function & $V(n)=v_f(1-\frac{n}{n_{jam}})^2$ [m/s]\\
Learning parameter &  $\omega=0.7$\\
Toll profile function & $Toll(t|A, \xi, \sigma)=A\times e^{\frac{-(t-\xi)^2}{2\sigma^2}}$\\
\hline
\end{tabular}
\end{table}

In this study, we assume the capacity of the single-reservoir network as 4500 travelers, using speed function in \cite{lamotte2018morning} and other parameters adopted in \cite{yildirimoglu2020demand} and \cite{liu2020managing}. We can derive the flow-maximizing accumulation value (or critical value) $n_{cr}=1500$ [travelers]. To test the performance of the pricing scheme under a high congestion scenario, the demand is set at 3700 travelers so that the peak accumulation value at the no toll equilibrium state exceeds the critical value (see Section \ref{section:d2d_converge} for details). Besides, we form the heterogeneity of travelers by drawing their trip lengths and schedule deviation penalties from three truncated Gaussian distributions shown in Table \ref{settings}.

\subsection{Day-to-day evolution process}\label{section:d2d_converge}
In this section, we illustrate that with and without the TODP scheme, the proposed day-to-day evolution can reach the equilibrium states where the departure time choices of all travelers keep unchanged. Besides, the perceived generalized cost vector should equal to the experienced cost vector at the equilibrium, i.e.,
$\bm{C}_{*}=\bm{c}_{*}$. Hence we compute the inconsistency between $\bm{C}_{d}$ and $\bm{c}_{d}$ as $\parallel \bm{C}_{d}-\bm{c}_{d}\parallel_1/N$ to indicate the degree of convergence.

\begin{figure}[tb]
\centerline{\includegraphics[width=0.5\textwidth]{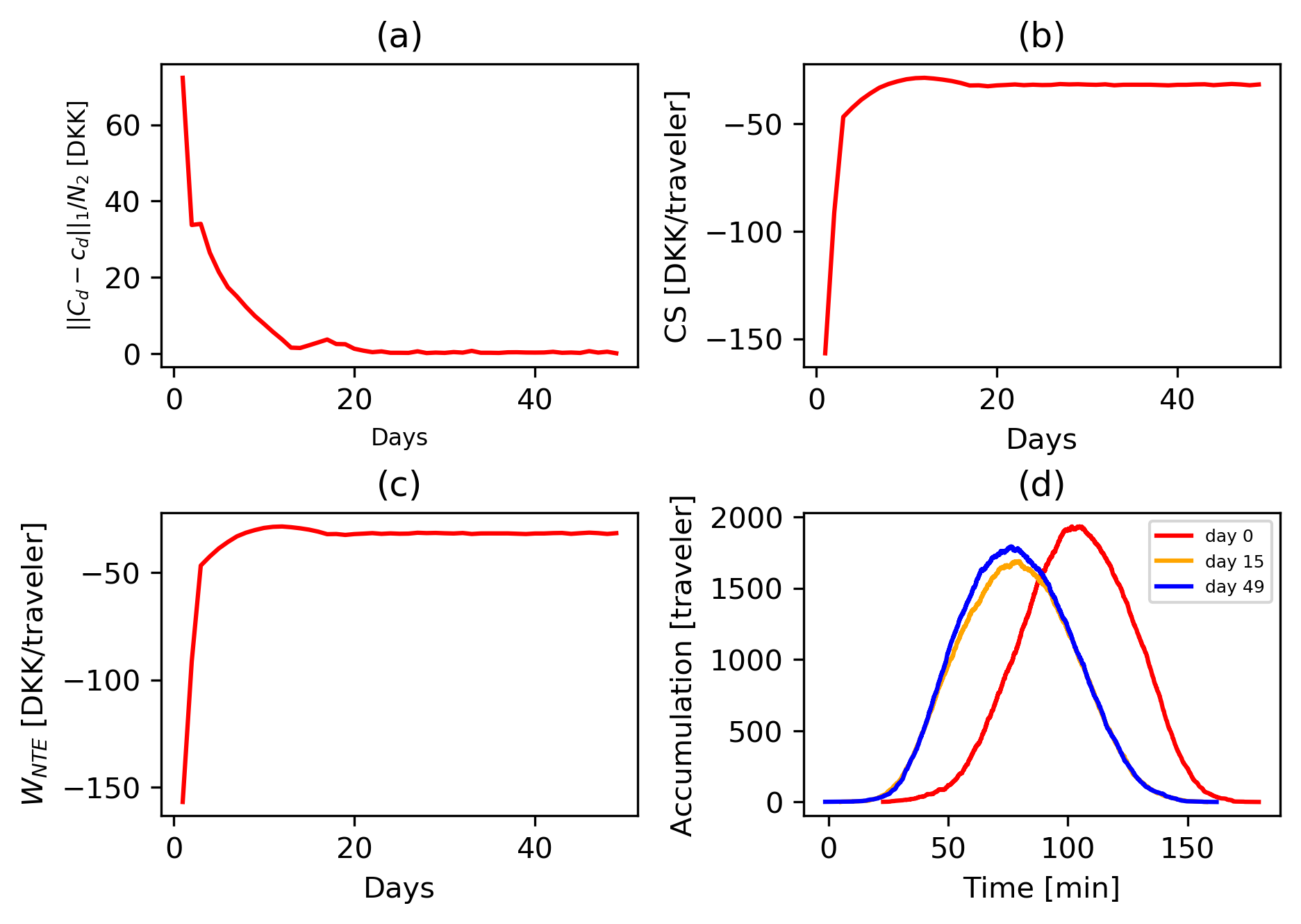}}
\caption{The evolution process of the no toll scenario.}
\label{NTE}
\end{figure}

Fig.~\ref{NTE}(a) presents the convergence process of the inconsistency in the no toll scenario, which comes close to 0 and stays stable after 25 days. Fig.~\ref{NTE}(b)-(c) show the evolution process of the average consumer surplus and social welfare per capita, respectively. Curves of these two plots are the same since the social welfare equals to the consumer surplus when there is no pricing. Fig.~\ref{NTE}(d) illustrate the within-day evolution of accumulation on different days, it can be found that the accumulation profile becomes stable after 25 days, and the peak accumulation exceeds the critical value $n_{cr}$. The observations from Fig.~\ref{NTE} imply that the equilibrium state of the day-to-day evolution is reached. Note that the inconsistency, consumer surplus and social welfare are plotted from day 1 as travelers do not have the perception of cost on day 0.

\begin{figure}[tb]
\centerline{\includegraphics[width=0.5\textwidth]{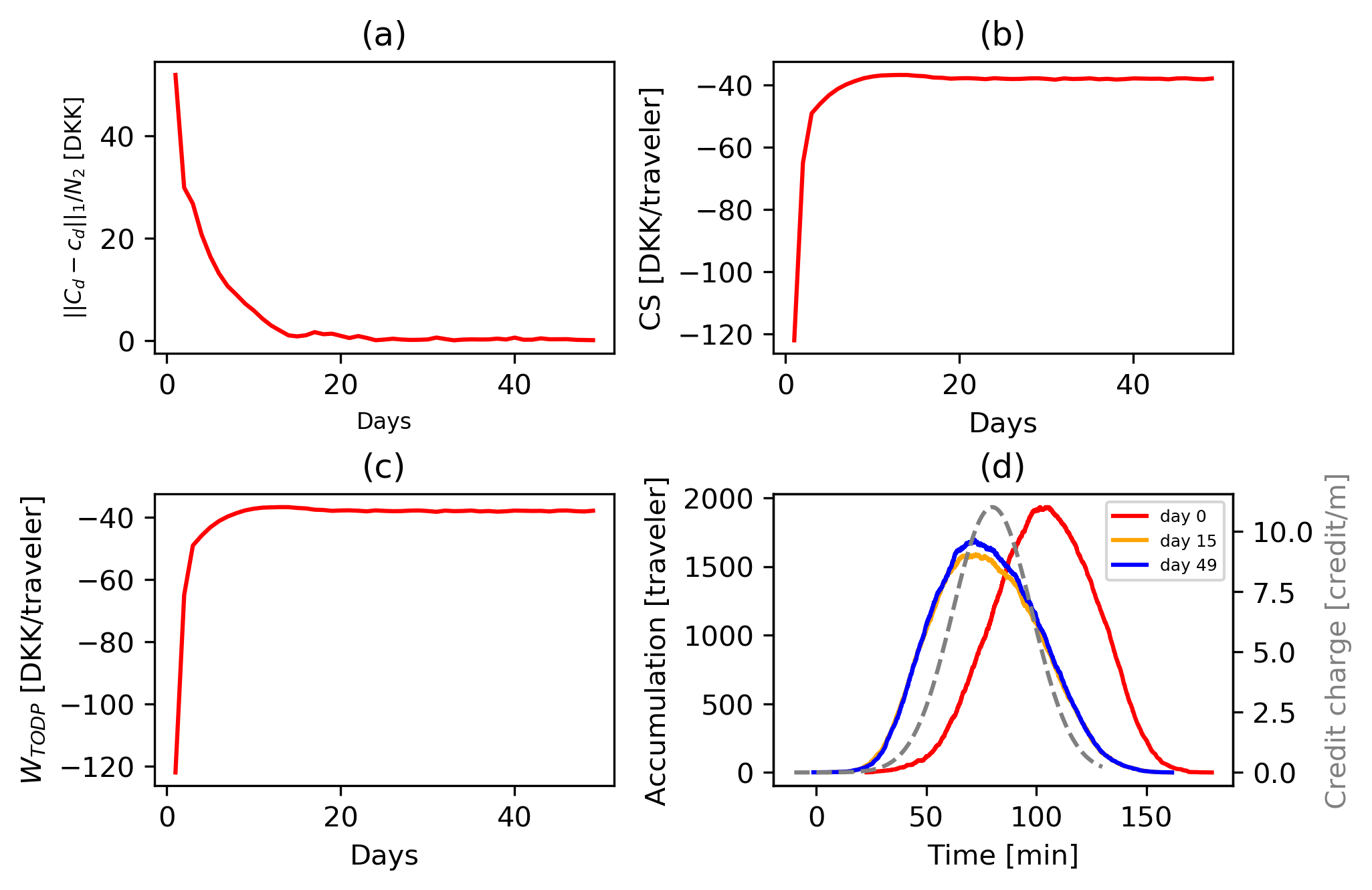}}
\caption{The evolution process under a TODP scheme.}
\label{TODP}
\end{figure}

To testify the convergence, we impose a TODP with parameters $A=11$, $\xi=80$ and $\sigma=18$ of day-to-day dynamics. In this case, the toll scheme is not optimized but arbitrarily given. As shown in Fig.~\ref{TODP}(a)-(c), the day-to-day evolution under a given TODP scheme also converges to the equilibrium after 25 days. In addition to accumulation profiles, Fig.~\ref{TODP}(d) also plots the TODP toll profile in a dashed grey line. It is found that travelers change their departure time to avoid the period with higher charge rate.

\subsection{Algorithm Performance}\label{section:performance}
In this section, we present the application of the simulation-based optimization framework to optimize the TODP parameters and evaluate the performance of BO. We first investigate the influence of the shape of TODP toll profile on social welfare by parametrizing the TODP with a mixture function, which vary from one (\textit{1G}) to six (\textit{6G}) Gaussian functions.

\begin{figure}[tb]
\centerline{\includegraphics[width=0.4\textwidth]{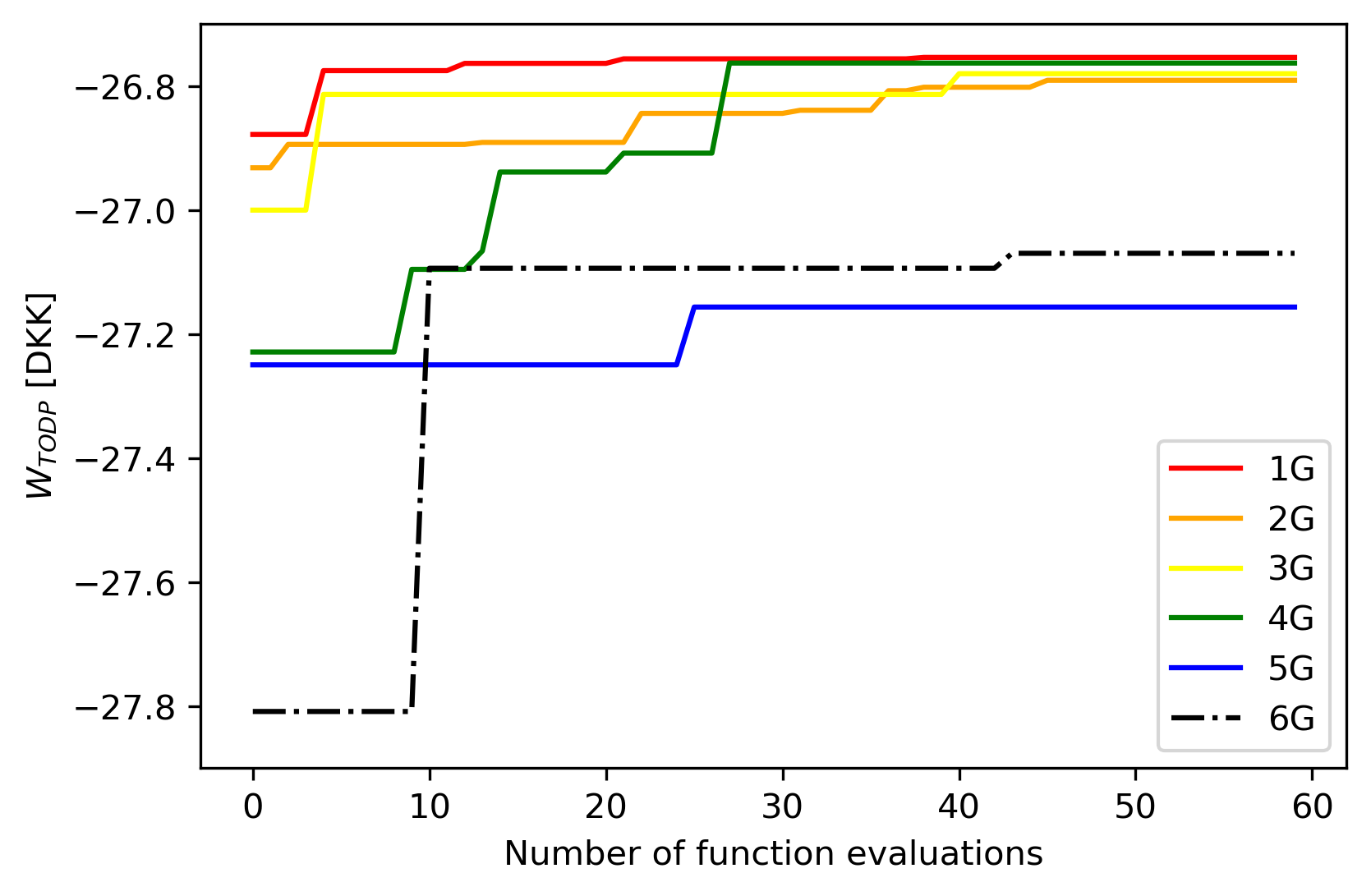}}
\caption{BO's performance with different dimensions of variables.}
\label{perform}
\end{figure}

In this test, the domains of the TODP function parameters (decision variables) are set as $A\in[4,30]$ (unit: DKK/m), $\xi\in[30,90]$ and $\sigma\in[10,50]$. To be fair, we generate 30 initial sampled sets of decision variables to fill the space for each case, and the number of function evaluations is fixed as 60. Fig.~\ref{perform} illustrates the evolution of the best social welfare with the number of function evaluations for BO using six different TODP functions. It can be found that BO has a relatively stable performance when the dimension of the decision variables is lower or equal to 12, and the obtained toll profiles have similar shapes. However, for the cases of \textit{5G} and \textit{6G} (15 and 18 parameters, respectively), standard BO fails to reach the maximal value obtained in lower dimension cases. The best solution found by BO is in the case of 1G, where $A=26.2$, $\xi=67.1$ and $\sigma=28.8$ with $W_{TODP}=-26.76$ [DKK]. 

Moreover, we adopt the BO dropout method to enhance the performance of BO for the case with high dimensions, e.g., for the case of \textit{6G} with 18 parameters. First, four cases with $d=1,3,5,7$ are tested to find out the effect of $d$. Since the dropout method chooses the decision variables in a random way, we propose two promising variable choosing strategies for further improvement with $d=5$: In strategy 1 (S1), we categorize the 18 variables into 3 groups by physical meanings, e.g., all the $A$s are in one group, and at least one variable of each group will be chosen in each iteration of BO; In strategy 2 (S2), we categorize the 18 variables into 6 groups by functions, e.g., $A_1$, $\xi_1$ and $\sigma_1$ are in a group, we randomly choose 5 out of the 6 groups and draw one variable in each chosen group. We run each case four times and summarize the average objective values and standard deviations in Table \ref{results}. The results demonstrate that all dropout cases reach higher objective values with smaller standard deviations compared to the standard BO. Besides, the proposed two strategies enhance the performance of the pure random dropout method with $d=5$, and strategy 2 even outperforms the random dropout method in all cases. This implies that performance improvement might be realized by developing case-specific proper rules on choosing variables. Nevertheless, more experiments are needed to investigate the influence of $d$ and dropout strategies in terms of uncertainty.


\begin{table}[b]
\centering
  \caption{Average Performances of Different Dropout Methods}\label{results}
  \vskip 0.2cm
  \begin{tabular}{lccccccc}
\hline\noalign{\smallskip}
Method & BO ($6G$)  & $d=1$ & $d=3$ & $d=5$ & $d=7$  & S1 & S2 \\
\noalign{\smallskip}\hline\noalign{\smallskip}
Mean & 27.00 &  26.81 & 26.82 & 26.85 & 26.85 &  26.82 & 26.79 \\
Std.dev & 0.17 & 0.04 & 0.07 & 0.05 & 0.08 &  0.05 & 0.01 \\

\hline
\end{tabular}
\end{table}

\subsection{Optimization results}\label{section:optimal}
\begin{figure}[tb]
\centerline{\includegraphics[width=0.45\textwidth]{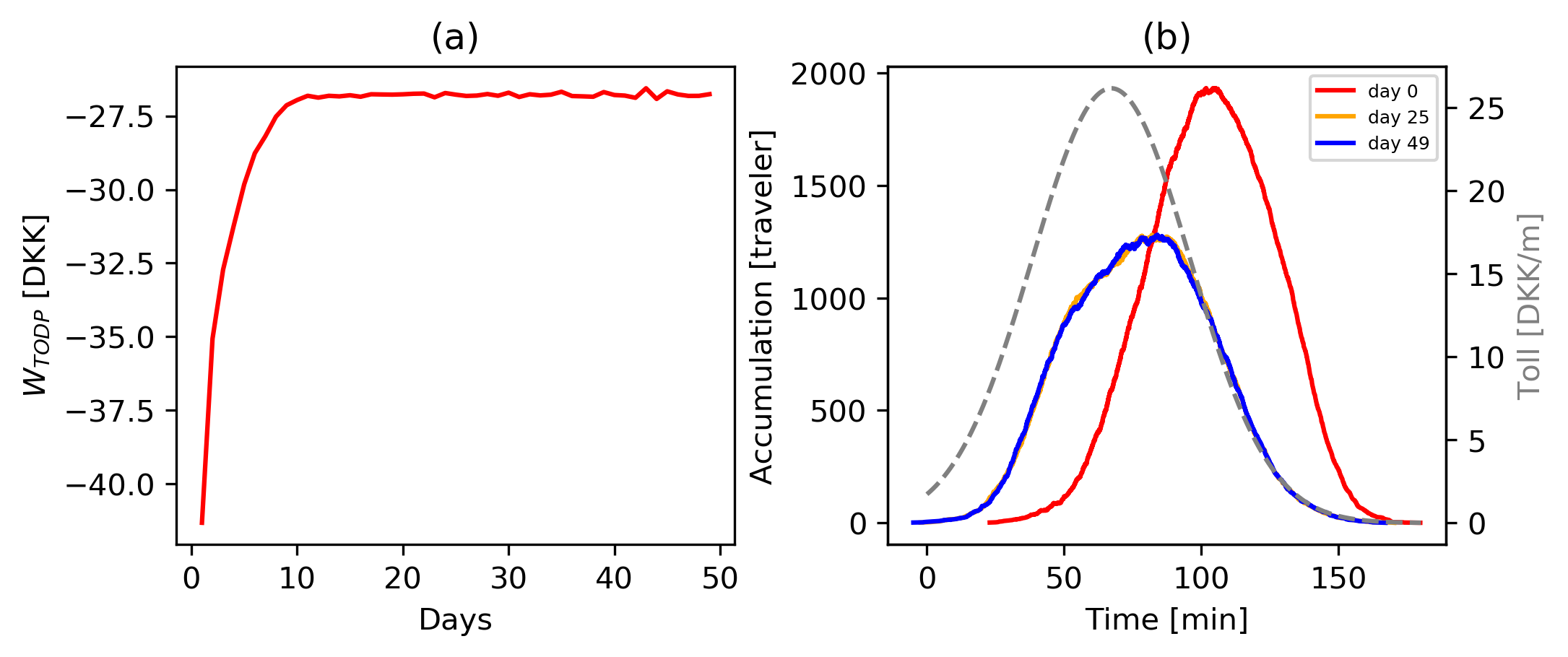}}
\caption{The evolution process under optimized TODP.}
\label{optimal}
\end{figure}

In this subsection we present the optimization results of the TODP scheme, wherein the toll profile is represented by a single Gaussian function. Fig. \ref{optimal}(a) illustrates the evolution of social welfare per capita and Fig. \ref{optimal}(b) plots the variations of accumulation and the optimized toll profile. Compared to the no toll case, the peak accumulation is reduced from 1792 to 1281 [traveler], leading to 16.0\% improvement in the social welfare per capita from -31.86 [DKK] to -26.76 [DKK]. Moreover, the average travel time cost is reduced by 3.4\% and the average schedule deviation cost is increased by 37.2\%.

\section{Conclusion}
This paper focuses on the dynamics of TODP, under heterogeneous demand and a trip-based MFD network simulation model, and presents a simulation-based BO framework for optimizing the pricing scheme in terms of social welfare.

The performed numerical results show that BO performs worse with higher dimensions in the decision variable space, as the needed number of function evaluations increases exponentially to find the optimum. When testing the dropout method proposed by \cite{ijcai2017-291}, we were able to reach higher dimension search, which is quite relevant for the problem at stake when implementing detailed pricing schemes. We then propose a model-informed approach to formulate problem-specific dropout strategies, and showcase how selecting related toll profile features during the dropout process outperforms the existing random dropout method in terms of optimization and stability. Such insights encourages the testing of our proposed method to other practical but possibly high-dimension TODP schemes, such as step tolls or multiple area networks. Finally, our optimized TODP is compared against the no-control strategy, showing clear benefits in terms of social welfare and network performance.

\section*{Acknowledgment}

This research was carried out under the NEMESYS project funded by the DTU-NTU Alliance.


\bibliographystyle{ieeetr}
\bibliography{ref.bib}

\end{document}